\documentstyle[preprint,prb,aps]{revtex}
\tighten
\begin{document}
\draft
\preprint{LA-UR-94-2590 {\bf /} MA/UC3M/11/94}

\title{Incoherent exciton trapping in self-similar aperiodic lattices}

\author{Francisco\ Dom\'{\i}nguez-Adame and Enrique\ Maci\'{a}$^*$}
\address{Departamento de F\'{\i}sica de Materiales,
Facultad de F\'{\i}sicas, Universidad Complutense,\\
E-28040 Madrid, Spain}

\author{Angel S\'{a}nchez}
\address{Theoretical Division and Center for Nonlinear Studies, Los
Alamos National Laboratory,\\ Los Alamos, New Mexico 87545\\ and\\
Escuela Polit\'{e}cnica Superior, Universidad Carlos
III de Madrid, C./ Butarque 15,\\ E-28911 Legan\'{e}s,
Madrid, Spain}

\maketitle

\begin{abstract}

Incoherent exciton dynamics in one-dimensional perfect lattices with
traps at sites arranged according to aperiodic deterministic sequences
is studied.  We focus our attention on Thue-Morse and Fibonacci systems
as canonical examples of self-similar aperiodic systems.  Solving
numerically the corresponding master equation we evaluate the survival
probability and the mean square displacement of an exciton initially
created at a single site.  Results are compared to systems of the same
size with the same concentration of traps randomly as well as
periodically distributed over the whole lattice.  Excitons progressively
extend over the lattice on increasing time and, in this sense, they act
as a probe of the particular arrangements of traps in each system
considered.  The analysis of the characteristic features of their time
decay indicates that exciton dynamics in self-similar aperiodic
arrangements of traps is quite close to that observed in periodic ones,
but differs significatively from that corresponding to random lattices.
We also report on characteristic features of exciton motion suggesting
that Fibonacci and Thue-Morse orderings might be clearly observed by
appropriate experimental measurements.  In the conclusions we comment on
the implications of our work on the way towards a unified theory of the
orderings of matter.

\end{abstract}


\narrowtext

\section{Introduction}

The interest in the study of the physical properties of elementary
excitations in one-dimensional (1D) self-similar aperiodic systems has
considerably grown during the last years.  Albeit these systems were
originally considered as somewhat intermediate between the periodic
(crystalline) and random (amorphous, glassy) orderings of matter, it has
been progressively realized that systems containing basic units arranged
according to the Fibonacci,\cite{Kohmoto,Das,Chakrabarti}
Thue-Morse,\cite{Riklund,Tamura} period-doubling \cite{Barache} or
Rudin-Shapiro \cite{Dulea} sequences display novel properties which are
not shared by the systems usually considered in condensed matter
physics.  In this way, we have recently provided strong evidence
supporting the idea that self-similar aperiodic systems reveal a new
kind of order, rather than representing a confuse mixture of periodic
order and randomness.\cite{PRB4,PRE} From a theoretical point of view
this line of reasoning stems from the fact that, in spite of Bloch
theorem being not longer valid for arbitrary aperiodic systems lacking
translational symmetry, there exist certain aperiodic systems which
still endow a significative degree of structural symmetry.  Between
them, those systems generated from the application of a substitution
sequence \cite{Bellissard} deserve an especial attention, for they
exhibit a characteristic {\em scale invariance symmetry} directly
related to their self-similar nature.

As experimental realizations of such systems become available in the
fields of quasicrystalline phase research and multilayered
heteroestructures technology, the interest in these aperiodically
ordered forms of matter goes beyond a mere conceptual interest.  For the
sake of illustration let us mention but the following examples.  In the
area of quasicrystalline matter the rapidly solidified alloys
Al$_{80}$Ni$_{14}$Si$_6$, Al$_{65}$Cu$_{20}$Mn$_{15}$,
Al$_{65}$Cu$_{20}$Co$_{15}$, which are crystalline in two orthogonal
directions and quasicrystalline along the other one,\cite{He} are worth
being mentioned.  On the other side, regarding multilayered structures,
after the first fabrication of Fibonacci \cite{Merlin85} and Thue-Morse
\cite{Merlin87} semiconductor superlattices by Merlin and co-workers,
the obtention of metallic \cite{Carlotti,MoV} and superconducting
quasiperiodic superlattices \cite{Cohn} and networks \cite{Nori} has
been reported.

It is well known that self-similar aperiodic systems, described by
tight-binding and Kroning-Penney models, possess singular continuous
energy spectra which are Cantor sets of zero Lebesgue measure.  This
point has been rigorously proven for Fibonacci,\cite{Fibo,Fibo2}
period-doubling, and Thue-Morse\cite{Thue} sequences and it has recently
been conjectured, on sound mathematical basis, that this spectral type
may be a common characteristic of all aperiodic systems obtained by the
application of a substitution sequence.\cite{Bovier94} In this sense, a
great deal of numerical analyses have shown that this kind of spectra
exhibit a highly fragmented structure, with a hierarchy of splitting
subbands displaying self-similar patterns and that the associated
(generalized) eigenstates behave in a very peculiar manner, referred to
as {\em critical}, characterized by dramatic spatial fluctuations and
becoming neither localized nor extended in the usual
sense.\cite{Severin,Kim,Angel,Oh,Zhong} In addition, realistic
estimations on the possibility of observing characteristic quasiperiodic
effects in more complex systems, like Fibonacci Si $\delta$-doped GaAs
superlattices, has been recently reported.\cite{PLA}

Hence, the question as to whether the peculiar structure of the energy
spectrum of self-similar aperiodic systems influences the transport
properties through the sample follows in a natural way.\cite{PRB4,Angel}
Most studies dealing with this topic up to date have been mainly
concerned in both electronic and phonon transport problems, which can be
treated, in a unified and simple mathematical scheme, within the
transfer-matrix approach.  Nevertheless, interesting results involving
other elementary quasiparticles, including plasmons,\cite{Hawrylak}
photons,\cite{Gellermann} excitons,\cite{OPTFIBO} and
polaritons,\cite{Albuquerque} in aperiodic lattices have been reported,
thus considerably widening the field of possible practical applications
of these systems.

In this work we will investigate {\em incoherent} exciton dynamics in 1D
self-similar aperiodic systems, considering the Fibonacci and Thue-Morse
sequences as canonical examples.  The main aim of this study is twofold.
In the first place we ascertain how self-similar order modifies exciton
dynamics in comparison with the dynamics associated to the long-range
disorder of random systems.  To this end we adopt the following
strategy: We solve numerically the master equation governing the exciton
motion and, from its solution, we obtain the survival probability and
the mean square displacement of excitons.  These parameters are then
compared to those found in random lattices of the same size with
identical fraction of traps.  In the second place we determine the
differences between exciton propagation through periodic chains and
exciton transport in aperiodic systems displaying quasiperiodic order
(Fibonacci) on one hand, and non-quasiperiodic order (Thue-Morse) on the
other hand.  In this way we are able to report on two interesting
results.  First, at least regarding exciton dynamics, self-similar
aperiodic lattices are more similar to periodic lattices than they are
to random ones and, secondly, exciton motion in quasiperiodic lattices
also differs from the corresponding motion in non-quasiperiodic chains.
In view of these results the notions of both {\em quasiperiodic order}
and {\em self-similar order}, as denoting new and different classes of
matter ordering, seem completely justified.\cite{PRE,OPTFIBO}

Finally, along with the possible applications of our results to the
physical systems indicated previously ---it is worth mentioning here
that for short and intermediate times 1D transport may be relevant for
three dimensional systems as well \cite{Scher}---, one of the most
appealing aspects of the present work is to show how time evolution of
quasiparticles (excitons in the present case) may be usefully employed
to determine structural features of lattices.  In particular, we
demonstrate that excitons, initially created at a single site, act as a
probe of the underlying structure as time evolves and the quasiparticle
interacts with larger and larger regions of the system via the combined
action of diffusion and trapping.

We will report on these issues according to the following scheme.  In
Sec.~II we describe our model and the physical magnitudes we will
compute in order to properly characterize exciton dynamics.  Section III
contains our main results concerning survival probabilities and mean
square displacements of incoherent excitons, along with the
corresponding interpretation of the obtained results.  Section IV
concludes the paper with a brief account on practical implications of
our results and comments on some general ideas concerning the notion of
{\em aperiodic order} stemming from our study.

\section{Model}

We consider excitations in a 1D lattice whose time evolution is
described by the following master equation for the probability $P_k(t)$
to find the exciton at site $k$ \cite{Alexander}
\begin{equation}
{d\over dt}P_k = W(P_{k+1}+P_{k-1}-2P_k)-G_kP_k,
\label{master}
\end{equation}
where $W>0$ is the intersite rate constant, which is assumed to be
independent of $k$ hereafter and $G_k=G$ is the trapping rate at site
$k$.  The quantity of interest in luminescence experiments is the
survival probability $n(t)$ defined as
\begin{equation}
n(t)= \sum_k\> P_k(t),
\label{survival}
\end{equation}
where the index $k$ runs over all lattice sites. Moreover, assuming that
the excitation is initially at site $k_0$ ($P_k(0)=\delta_{kk_0}$), we
can also calculate the mean square displacement of the excitation (which
is related to the diffusion coefficient \cite{Alexander}) as follows
\begin{equation}
R^2(t)= \sum_k\> (k-k_0)^2 P_k(t),
\label{square}
\end{equation}
where the lattice spacing is taken to be unity hereafter.  These two
functions characterize the dynamics of excitons in the lattice under the
combined action of diffusion and trapping. Thus, it is known that, in
infinite lattices without traps ($G_k=0$), the survival probability is
conserved ($n(t)=1$) and the mean square displacement increases
linearlywith time \cite{Bartolo} ($R^2(t)=2Dt$, $D$ being the diffusion
coefficient).

In what follows we consider that $G_k$ can only take on two values,
$G_A=0$ and $G_B=G>0$, that is, only sites $B$ are able to trap
excitons.  We will arrange sites $A$ and $B$ according to the Thue-Morse
sequence, the Fibonacci sequence, at random, or periodically, depending
on the particular kind of lattice we are interested in.  For convenience
we define $c$ as the ratio between the number of traps and the total
number of sites in the considered lattice $N$.  Deterministic aperiodic
sequences can be generated by simple substitution rules.  Thus, we have
$A\rightarrow AB$, $B\rightarrow BA$ for the Thue-Morse sequence and
$A\rightarrow AB$, $B\rightarrow A$ for the Fibonacci one.  Finite,
self-similar lattices are obtained in this way by $l$ successive
applications of the substitution rule.  The {\em l\/}th generation
lattice has $2^l$ elements for the Thue-Morse lattice (TML) and $F_l$
elements for the Fibonacci lattice (FL), where $F_l$ denotes the
Fibonacci numbers.  Such numbers are generated from the recurrence
relationship $F_l=F_{l-1}+F_{l-2}$ with $F_0=F_1=1$; as $l$ increases
the ratio $F_{l-1}/F_l$ converges toward $\tau = (\sqrt{5}-1)/2 =
0.618\ldots$, which is known as the inverse golden mean.  Therefore,
sites are arranged according to the sequence
$A\,B\,B\,A\,B\,A\,A\,B\ldots$ in the TML, and
$A\,B\,A\,A\,B\,A\,B\,A\ldots$ in the FL. The value of $c$ is strictly
equal to $0.5$ for any generation of the TML. On the contrary, the value
of $c$ depends on the particular generation of the FL, but for large
enough systems one has $c\sim 1-\tau=0.3819\ldots$.  Disordered lattices
are obtained by placing traps (sites $B$) at random over the lattice
maintaining fixed the concentration of traps $c$.  Finally, we consider
in this work periodic lattices of two types.  One of them is set with
$c=0.5$ and traps placed at sites with even index.  This periodic
lattice will be compared to the TML. The other type is obtained from a
periodic superposition of unit cells of the form
$A\,B\,A\,A\,B\,A\,B\,A$, which is nothing but the fifth order
approximant to the Fibonacci sequence.  The concentration of traps is
$c=0.375$ for this periodic arrangement, a value rather close to the
value $1-\tau$ corresponding to infinite FLs.

\section{Numerical results and discussions}

We have numerically solved the master equation (\ref{master}) for TML,
random and periodic lattices of $N=2^{10}=1024$ units and for FL, random
and periodic lattices of $N=F_{15}=987$ units using an implicit
(Crank-Nicholson) integration scheme.  To avoid free ends effects,
spatial periodic boundary conditions are introduced, so that the
detailed balance required by Eq.\ (\ref{master}) is preserved.  The
initial condition for the exciton motion is $P_k(0)=\delta_{kk_0}$, with
$k_0=500$ ($k_0=494$) for lattices with $N=1024$ ($N=987$) sites, that
is, we will assume that the exciton is created, for instance by a pulsed
excitation, roughly at the middle of the lattice.  Trapping rate $G$
will be measured in units of $W$ whereas time will be expressed in units
of $W^{-1}$.  The maximum integration time and the integration step are
$250$ and $5\times 10^{-4}$, respectively.  Smaller time steps led to
similar results.  Since we are mainly interested in the effects due to
particular arrangements of traps rather than in a detailed description
of the influence that the different parameters have in the incoherent
motion of excitations, we will fix the values of $W$ and $G$ henceafter.
Thus we have set $W=1$ and $G=0.2$ as representative values.  For
disordered lattices a series of random distribution of traps was
generated for a given trap concentration, and ensembles comprising a
number of realizations varying from $50$ to $200$ were averaged to check
the convergence of the computed mean values.  Since convergence was
always satisfactory between all the ensembles, we present the results
corresponding to $50$ averages.

The obtained results for the mean square displacement and survival
probability of excitons propagating through the TML are presented in
Figs.~\ref{fig1}(a) and \ref{fig1}(b), respectively, along with the
corresponding results for random and periodic lattices with a trap
concentration of $c=0.5$.  Analogous magnitudes describing the motion of
incoherent excitons through the FL and related random and periodic
lattices are shown in Figs.~\ref{fig2}(a) and \ref{fig2}(b).  Let us
consider, in the first place, the behavior of the mean square
displacement of incoherent excitons through these systems.  In all cases
it becomes apparent that the time evolution of $R^2(t)$ arises from the
competition between two different processes, namely diffusion (the
exciton is transferred from site to site, starting at $k_0$) and
trapping (the exciton progressively decays in time since possible
detrapping processes are not considered in our model).  At short times
the first mechanism dominates because of the exciton is still close to
the initial position and, consequently, there exist small chances to be
trapped.  As time elapses, the probability of trapping also increases
since the exciton can be found in a larger segment of the lattice.  This
competition gives rise to the occurrence of a well defined maximum in
$R^2(t)$, whose position depends not only on the concentration of traps
but mainly on the spatial distribution of these traps.  In addition to
this quite general behavior we observe significant differences between
the exciton behavior in quasiperiodic (Fibonacci) and non-quasiperiodic
(Thue-Morse) aperiodic lattices.  In fact, the mean square displacement
of an exciton propagating through a TML essentially {\em coincides} with
that corresponding to the case of the periodic lattice over the entire
time interval we have considered.  Moreover, the $R^2(t)$ curve
describing the exciton motion in the random lattice appreciably differs
from both the TML and periodic corresponding curves.  On the contrary,
the mean square displacement of excitons in the FL cannot be easily
compared with that of excitons moving in neither periodic nor random
lattices at short times but, as time increases, exciton motion in FLs
progressively resembles that taking place in the periodic lattice
approximant.

Now, we turn our attention to the evolution of the survival probability.
It is well known that, for any periodic distribution of traps, the
behavior of the survival probability is simply exponential in time, and
is given by the expression $n(t)= \exp(-cGt)$, whereas in random
lattices presents a more complex and non-exponential dependence on time.
Keeping this fact in mind, the interpretation of Figs.~\ref{fig1}(b) and
\ref{fig2}(b) is straightforward.  The rate of trapping of incoherent
excitons in both the TML and FL is very similar to that of the
corresponding periodic lattices with the same fraction of traps and
quite different from that associated with the corresponding random
lattices.  Therefore, from this point of view, self-similar aperiodic
systems behave as periodic ones in a very close manner.  In particular
we note that not only an exponential decay rate for both kind of
aperiodic systems is observed, but the slope of the corresponding
survival probabilities fits the value prescribed by the trap
concentration $c$ appearing in the general expression for periodic
systems.  Finally, note that the decay rate in random lattices is much
slower than in the other lattices (periodic and aperiodic).

\section{Conclusions}

{}From the comparison of the mean square displacement and survival
probability plots for the lattices considered in this work, several
conclusions can be drawn.  In the first place we point out that excitons
propagating through self-similar aperiodic lattices behave in a very
similar way as they do in periodic 1D systems.  We wish to stress, at
this respect, that we are not merely saying that excitonic behavior
resembles that corresponding to excitons moving in periodic lattices,
but stating that exciton dynamics in self-similar chains exhibits an
time evolution completely different from that they show in random
systems.  A second important result emerging from our numerical
simulations is that the exciton dynamics in the TML significatively
differs from that recorded in the FL at short times.  This can be easily
seen by comparing the corresponding mean square displacement curves.
The justification for this effect can be accounted for starting from the
following picture: As time evolves the exciton progressively extends
over the lattice and, in this sense, it acts as a probe indicating the
rate of trapping associated to the particular arrangement of traps of
the underlying structure.  In this way, the shape of the $R^2(t)$ curve
can be interpreted in a topological sense.  As it has been explained
previously the characteristic maximum of this curve indicates a cutoff
between two different transport regimes in the system.  At short times
we have classical diffusion through the lattice, meanwhile at longer
times the effects of trapping become dominant.  Fig.~\ref{fig1}(a)
indicates that excitons propagate through the TML as they will do
through a periodic lattice having the same trap concentration in both
regimes.  Hence, non-quasiperiodic order associated to the Thue-Morse
sequence has no relevant effects on exciton dynamics and, as long as
transport properties are concerned, Thue-Morse and binary periodic
arrangements with a trap concentration $c=0.5$ are completely
equivalent.  On the contrary, the shape of the $R^2(t)$ curve, shown in
Fig.~\ref{fig2}(a), clearly reveals that quasiperiodic order has a
profound effect on the excitonic diffusion transport regime.  In fact,
we see that diffusion of excitons in the FL is considerably lower than
that taking place in both periodic and random lattices at the same
times.  This result, along with the fact that the area determined from
the expression $\int_0^{\infty}R^2(t)\, dt$ is also smaller than the
corresponding values for the other lattices, led us to the conclusion
that trapping processes are {\em more efficient} for quasiperiodic
arrangements than they are for other possible orderings, including both
periodic (crystalline) and random (glassy) structures.  This interesting
result may be of relevance from an experimental point of view, since by
properly measuring the value of diffusion constants in aperiodic
lattices, we should be able to estimate the kind of underlying
topological order they present.

To conclude we wish to comment on some general ideas concerning the
notion of {\em aperiodic order} which stem from our study.  The results
reported on in this work provide substantial support to the view,
previously put forward from our study of electronic transport in
quasiperiodic systems,\cite{PRB4} that aperiodic systems cannot be
regarded, in general, as systems endowed with an intermediate degree of
structural disorder.  On the contrary, certain classes of aperiodic
systems, like those arranged according to the Fibonacci and Thue-Morse
sequences, are notable representatives of highly ordered systems,
actually displaying, in some particular aspects, a higher level of order
than usual periodic systems are able to do.\cite{PRE} In this way we
realize that it might well be that the long-standing creed of periodic
order as a canonical prototype of perfect order should be reconsidered.

A further step in this conceptual progress from the notion of randomness
to that of perfect order comes from the progressive realization that not
all self-similar aperiodic orderings can be put on the same footing.  In
fact, when comparing quasiperiodic and non-quasiperiodic arrangements by
means of different criteria we usually find that, just depending on the
adopted criteria, one kind of system seems to be more or less ordered
than the other one.  In this sense, it has been claimed, on the basis of
the electronic wavefunctions behavior, that the TMLs might be regarded
as providing a link between the FLs and the periodic lattices, and that
the study of non-quasiperiodic aperiodic systems is a promising way
towards a unified theory of ordered systems, able to encompass periodic
as well as aperiodic orderings of matter.\cite{Riklund87} This earlier
suggestion can be phrased in a novel fashion by introducing the concept
of {\em hierarchies of order\/}.\cite{OPTFIBO} By this we mean that
rather than thinking of different kinds of order, classified into
separated categories ranging from those more periodic to those more
random, as it has been the usual procedure in a number of recent
works,\cite{Roy,Ryu,Huang} it may be more fruitful to separate different
kinds of order in a qualitative way by grading them according to a well
established criterion.  Since only two kinds of binary aperiodic
lattices have been extensively studied to the date, it is hard to
propose such a criterion on a sound basis.  Nevertheless, it may be
confidently conjectured that it will ultimately have to do with the
self-similar properties of aperiodic deterministic lattices generated by
the application of substitution sequences.  Since self-similarity endows
these structures with scale invariance symmetry properties, we feel that
a systematic analysis of this kind of symmetry on rigorous mathematical
foundations is required before a proper understanding of aperiodic order
can definitively be attained.

\acknowledgments

This work is partially supported by Universidad Complutense through
project PR161/93-4811.  A.\ S.\ is partially supported by DGICyT (Spain)
grant PB92-0248, by MEC (Spain)/{}Fulbright, and by the European Union
Network ERBCHRXCT930413.  Work at Los Alamos is performed under the
auspices of the U.S.\ D.o.E.

\begin{figure}
\caption{(a) Mean square displacement and (b) logarithm of the survival
probability of excitons as a function of time for lattices of $N=1024$
sites with $c=0.5$.  Results correspond to Thue-Morse (solid lines),
periodic (long-dashed lines), and random (short-dashed lineas)
arrangements of traps.}
\label{fig1}
\end{figure}

\begin{figure}
\caption{(a) Mean square displacement and (b) logarithm of the survival
probability of excitons as a function of time for lattices of $N=987$
sites with $c=0.382$.  Results correspond to Fibonacci (solid lines),
periodic (long-dashed lines), and random (short-dashed lineas)
arrangements of traps.}
\label{fig2}
\end{figure}

\end{document}